Running head: Deterministic parallel communicating Watson-Crick automata systems
Title: Deterministic parallel communicating Watson-Crick automata systems
Authors: Kingshuk Chatterjee[1] ,Kumar Sankar Ray(corresponding author)[2]

Affiliations:
[1]Electronics and Communication Sciences Unit, Indian Statistical Institute, Kolkata-108

[2]Professor, Electronics and Communication Sciences Unit, Indian Statistical Institute, Kolkata-108

Address: Electronics and Communication Sciences Unit, Indian Statistical Institute, Kolkata-108.

Telephone Number: +918981074174

Fax Number:033-25776680

Email:ksray@isical.ac.in

# Deterministic parallel communicating Watson-Crick automata systems

Kingshuk Chatterjee,[1] Kumar Sankar Ray[2]

*Electronics and Communication Science unit, ISI, Kolkata.*

[1]kingshukchaterjee@gmail.com [2]ksray@isical.ac.in

*Abstract:* In this paper, we have introduced the deterministic variant of parallel communicating Watson-Crick automata systems. We show that similar to the non-deterministic version, the deterministic version can also recognise some non-regular uniletter languages. We further establish that strongly deterministic Watson-Crick automata systems and deterministic Watson-Crick automata system are incomparable in terms of their computational ability. We have also compared the computational ability of our system with multihead finite automata and parallel communicating finite automata systems.

## I. INTRODUCTION

Parallel communicating automata systems were introduced by Martin-Vide et.al [1]. The system consists of many finite automata communicating with states. They also established that the computational power of such a system is equivalent to non-deterministic finite automata with multiple heads and if the components of the system are deterministic in nature then the computational power is same as that of a deterministic multihead finite automata.

Watson-Crick automata are finite automata having two independent heads working on double strands where the characters on the corresponding positions of the two strands are connected by a complementarity relation similar to the Watson-Crick complementarity relation. The movement of the heads although independent of each other is controlled by a single state. Watson-Crick automata were introduced by Păun et.al.[2], its deterministic variants were introduced by Czeizler et.al. [3]. Work on state complexity of Watson-Crick automata is discussed in [4] and [5].

Parallel Communicating Watson-Crick automata systems (PCWKS) were introduced in [6] and further investigated in [7]. A parallel communicating Watson-Crick automata system [6] consists of several Watson-Crick automata working synchronously, each on its own input tape, and communicating on request. Special query states are provided, each of them pointing to exactly one component of the system. When a component i of the system reaches a query state $K_j$, the current state of the component j is communicated to i and the computation continues. There are two important classifications of parallel communicating systems. An automata system is called centralized if only one component, the master, may introduce query states, and non-centralized otherwise. An automata system is called returning if after communicating, a component resumes the computation from its initial state, and non-returning if it remains in its current state. Every component of parallel communicating Watson-Crick automata system has its own double-stranded tape; the input is the same on all of them. At the beginning, all components are in their initial states and start parsing synchronously the input from left to right. An input is accepted by the system if all components are in final states and they completely parse the tape. Moreover, if one of the components stops before the others, the system halts and rejects the input. Hence, in order to accept, the components either finish at the same time or wait for each other at the end of the computation. Czeizler et.al. [7] showed that with non-injective complementarity relation PCWKS can accept non-regular uniletter languages. In this paper we introduce the deterministic variant of such a parallel communicating Watson-Crick automata system and further show that similar to the non-deterministic system the deterministic system with non-injective complementarity relation can accept a non-regular uniletter language. We show that a strongly deterministic parallel communicating system with n degrees has the same computational power as a deterministic multi head finite automata with 2n heads. We further show that strongly deterministic parallel communicating Watson-Crick automata and deterministic Watson-Crick automata are incomparable in terms of language recognized which is also true for non-deterministic Watson-Crick automata.

In this paper, we give a general description of non-deterministic and deterministic Watson-Crick automata and its different variants in section 2,3 and 4. In the following section we state the rules governing parallel communicating Watson-Crick automata systems. In section 6, we introduce the rules of deterministic parallel communicating Watson-Crick automata systems. We further discuss the computational complexity of a such a system in Section 7. We conclude our work in Section 8.

## II. BASIC TERMINOLOGY

The symbol V denotes a finite alphabet. The set of all finite words over V is denoted by $V^*$, which includes the empty word λ. The symbol $V^+ = V^* - \{λ\}$ denotes the set of all non-empty words over the alphabet V. For $w \in V^*$, the length of w is denoted by |w|. Let $u \in V^*$ and $v \in V^*$ be two words and if there is some word $x \in V^*$, such that v=ux, then u is a prefix of v, denoted by

u ≤ v. Two words, u and v are prefix comparable denoted by u~$_p$v if u is a prefix of v or vice versa.

A Watson-Crick automaton is a 6-tuple of the form M=(V,ρ,Q,q$_0$,F,δ) where V is an alphabet set, set of states is denoted by Q, ρ ⊆ V×V is the complementarity relation similar to Watson-Crick complementarity relation, q$_0$ is the initial state and F⊆Q is the set of final states. The function δ contains a finite number of transition rules of the form $q\binom{w_1}{w_2}\to q'$, which denotes that the machine in state q parses w$_1$ in upper strand and w$_2$ in lower strand and goes to state q' where w$_1$, w$_2$∈V$^*$. The symbol $\begin{bmatrix}w_1\\w_2\end{bmatrix}$ is different from $\binom{w_1}{w_2}$. While $\binom{w_1}{w_2}$ is just a pair of strings written in that form instead of (w$_1$,w$_2$), the symbol $\begin{bmatrix}w_1\\w_2\end{bmatrix}$ denotes that the two strands are of same length i.e. |w$_1$|=|w$_2$| and the corresponding symbols in two strands are complementarity in the sense given by the relation ρ. The symbol $\begin{bmatrix}V\\V\end{bmatrix}_\rho = \{\begin{bmatrix}a\\b\end{bmatrix} \mid a, b \in V, (a,b) \in \rho\}$ and $WK_\rho(V) = \begin{bmatrix}V\\V\end{bmatrix}_\rho^*$ denotes the Watson-Crick domain associated with V and ρ.

A transition in a Watson-Crick finite automaton can be defined as follows:

For $\binom{x_1}{x_2},\binom{u_1}{u_2},\binom{w_1}{w_2} \in \binom{V^*}{V^*}$ such that $\begin{bmatrix}x_1u_1w_1\\x_2u_2w_2\end{bmatrix} \in WK_\rho(V)$ and $q,q' \in Q$, $\binom{x_1}{x_2}q\binom{u_1}{u_2}\binom{w_1}{w_2} \Rightarrow \binom{x_1}{x_2}\binom{u_1}{u_2}q'\binom{w_1}{w_2}$ iff there is transition rule $q\binom{u_1}{u_2}\to q'$ in δ and $\stackrel{*}{\Rightarrow}$ denotes the transitive and reflexive closure of ⇒. The language accepted by a Watson-Crick automaton M is L(M)={w$_1$∈V$^*$|$q_0\begin{bmatrix}w_1\\w_2\end{bmatrix} \stackrel{*}{\Rightarrow} q\begin{bmatrix}\lambda\\\lambda\end{bmatrix}$, with q ∈ F, w$_2$∈V$^*$, $\begin{bmatrix}w_1\\w_2\end{bmatrix}$ ∈WK$_\rho$(V)}.

### III. SUBCLASSES OF NON-DETERMINISTIC WATSON-CRICK AUTOMATA

Depending on the type of states and transition rules there are four types or subclasses of Watson-Crick automata. A Watson-Crick automaton M=(V,ρ,Q,q$_0$,F, δ) is

1) stateless( NWK ): If it has only one state, i.e. Q=F={ q$_0$ };
2) all-final( FWK ): If all the states are final, i.e. Q=F;
3) simple( SWK ): If at each step the automaton reads either from the upper strand or from the lower strand, i.e. for any transition rule $q\binom{w_1}{w_2}\to q'$, either w$_1$= λ or w$_2$= λ;
4) 1-limlited( 1-limited WK ): If for any transition rule $q\binom{w_1}{w_2}\to q'$, we have |w$_1$w$_2$|=1.

### IV. DETERMINISTIC WATSON-CRICK AUTOMATA AND THEIR SUBCLASSES

The notion of determinism in Watson-Crick automata and a discussion on its complexity were first considered in [3]. In [3] different notions of determinism were suggested as follows:

1) weakly deterministic Watson-Crick automata(WDWK): Watson-Crick automaton is weakly deterministic if in every configuration that can occur in some computation of the automaton, there is a unique possibility to continue the computation, i.e. at every step of the automaton there is at most one way to carry on the computation.
2) deterministic Watson-Crick automata(DWK): deterministic Watson-Crick automaton is Watson-Crick automaton for which if there are two transition rules of the form $q\binom{u}{v}\to q'$ and $q\binom{u'}{v'}\to q''$ then u≁$_p$u' or v≁$_p$v'.
3) strongly deterministic Watson-Crick automata(SDWK): strongly deterministic Watson-Crick automaton is a deterministic Watson-Crick automaton where the Watson-Crick complementarity relation is injective.

Similar to non-deterministic Watson-Crick automata, deterministic Watson-Crick automata can be stateless (NDWK), all final (FDWK), simple (SiDWK) and 1-limited (1-limited DWK).

### V. PARALLEL COMMUNICATING WATSON-CRICK AUTOMATA SYSTEM

A parallel communicating Watson-Crick automata system of degree *n*, denoted by PCWK(n), is a *(n + 3)*-tuple

$\mathcal{A} = (V, \rho, A_1, A_2, \ldots, A_n, K)$,

where

- *V* is the input alphabet;
- *ρ* is the complementarity relation;
- $A_i = (V, \rho, Q_i, q_i, F_i, \delta_i)$, $1 \le i \le n$, are Watson-Crick finite automata, where the sets $Q_i$ are not necessarily disjoint;

- $K = \{K_1, K_2, \ldots, K_n\} \subseteq \bigcup_{i=1}^{n} Q_i$ is the set of query states.

The automata $A_1, A_2, \ldots, A_n$ are called the *components* of the system A. Note that any Watson-Crick finite automaton is a parallel communicating Watson-Crick automata system of degree 1.

A configuration of a parallel communicating Watson-Crick automata system is a 2n-tuple $(s_1, \binom{u_1}{v_1}, s_2, \binom{u_2}{v_2}, \ldots, s_n, \binom{u_n}{v_n})$ where $s_i$ is the current state of the component $i$ and $\binom{u_i}{v_i}$ is the part of the input word which has not been read yet by the component $i$, for all $1 \leq i \leq n$. We define a binary relation $\vdash$ on the set of all configurations by setting

$$(s_1, \binom{u_1}{v_1}, s_2, \binom{u_2}{v_2}, \ldots, s_n, \binom{u_n}{v_n}) \vdash (r_1, \binom{u'_1}{v'_1}, r_2, \binom{u'_2}{v'_2}, \ldots, r_n, \binom{u'_n}{v'_n})$$

if and only if one of the following two conditions holds:

- $K \cap \{s_1, s_2, \ldots, s_n\} = \emptyset$, $\binom{u_i}{v_i} = \binom{x_i}{y_i}\binom{u'_i}{v'_i}$, and $r_i \in \delta_i(s_i, \binom{x_i}{y_i})$, $1 \leq i \leq n$;

- for all $1 \leq i \leq n$ such that $s_i = K_{j_i}$ and $s_{j_i} \notin K$ we have $r_i = s_{j_i}$, whereas for all the other $1 \leq \ell \leq n$ we have $r_\ell = s_\ell$. In this case $\binom{u'_i}{v'_i} = \binom{u_i}{v_i}$, for all $1 \leq i \leq n$.

If we denote by $\vdash^*$ the reflexive and transitive closure of $\vdash$, then the language recognized by a PCWKS is defined as:

$$L(\mathcal{A}) = \{w_1 \in V^* \mid (q_1, \begin{bmatrix}w_1\\w_2\end{bmatrix}, q_2, \begin{bmatrix}w_1\\w_2\end{bmatrix}, \ldots, q_n, \begin{bmatrix}w_1\\w_2\end{bmatrix}) \vdash^* (s_1, \begin{bmatrix}\lambda\\\lambda\end{bmatrix}, s_2, \begin{bmatrix}\lambda\\\lambda\end{bmatrix}, \ldots, s_n, \begin{bmatrix}\lambda\\\lambda\end{bmatrix}), s_i \in F_i, 1 \leq i \leq n\}.$$

Intuitively, the language accepted by such a system consists of all words $w_1$ such that in every component we reach a final state after reading all input $\begin{bmatrix}w_1\\w_2\end{bmatrix}$. Moreover, if one of the components stops before the others, the system halts and rejects the input. The above definition of parallel communicating Watson-Crick automata is in [6].

### VI. DETERMINISTIC PARALLEL COMMUNICATING WATSON-CRICK AUTOMATA SYSTEM

The notion of determinism in deterministic parallel communicating Watson-Crick automata is as follows.

1) weakly deterministic parallel communicating Watson-Crick automata system(WDPCWKS): a parallel communicating Watson-Crick automata system is weakly deterministic if every component in the system is a weakly deterministic Watson-Crick automaton.

2) deterministic parallel communicating Watson-Crick automata system(DPCWKS): a parallel communicating Watson-Crick automata system is deterministic if every component in the system is a deterministic Watson-Crick automaton.

3) strongly deterministic parallel communicating Watson-Crick automata system(SDPCWKS): a parallel communicating Watson-Crick automata system is strongly deterministic if every component in the system is a strongly deterministic Watson-Crick automaton.

### VII. COMPLEXITY OF DETERMINISTIC PARALLEL COMMUNICATING WATSON-CRICK AUTOMATA SYSTEM

In this section, we discuss the computational complexity of deterministic parallel communicating Watson-Crick automata systems.

**Theorem 1:** Every deterministic parallel communicating Watson-Crick automata system is equivalent with a system where in every component it has only rules of the form $q_i\binom{w_1}{w_2} \rightarrow q_j$ with $|w1w2| \leq 1$.

Proof: The following proof is similar to the proof described in [6]. We prove the above result for deterministic parallel communicating Watson-Crick automata as follows;

Let $A = (V, \rho, A_1, \ldots, A_n, K)$ be a deterministic parallel communicating Watson-Crick automata system with $n$ components, where $A_i = (V, \rho, Q_i, q_i, F_i, \delta_i)$ for all $1 \leq i \leq n$ are deterministic Watson-Crick automata. Let us first order the transitions from all components in a particular order and let its position in the order be its index. *We* define the constant $m = max\{|w1| + |w2| \mid q_i\binom{w_1}{w_2} \rightarrow q_j$ is a production in one of the components. The deterministic parallel communicating Watson-Crick automata system $A'(V, \rho, A'_1, \ldots, A'_n, K)$ which accepts the same language as A and has rules only of the form $q_i\binom{w_1}{w_2} \rightarrow q_j$ *with* $|w1w2| \leq 1$ in its components is constructed in the following manner:

Let $q_i\binom{w_1 w_2 w_3 \ldots w_k}{w'_1 w'_2 w'_3 \ldots w'_k} \rightarrow q_j$ with $w_1, w_2, w_3, \ldots w_k, w'_1, w'_2 w'_3 \ldots w'_k \in V$ be transition rule from $A_i$, indexed with the unique label j. Then, in $A'_i$ m new states $r_1^j, r_2^j, r_3^j, r_4^j, \ldots r_m^j$ and the following transitions are introduced.

$$q_i\binom{w_1}{\lambda} \to r_1^j \ldots\ldots\ldots\ldots\ldots r_{k-1}^j\binom{w_k}{\lambda} \to r_k^j,$$

$$r_k^j\binom{\lambda}{w'_1} \to r_{k+1}^j, \ldots\ldots\ldots\ldots r_{k+k'-1}^j\binom{\lambda}{w'_{k'}} \to r_{k+k'}^j,$$

$$r_{k+k'}^j\binom{\lambda}{\lambda} \to r_{k+k'+1}^j \ldots\ldots\ldots\ldots r_m^j\binom{\lambda}{\lambda} \to q_j.$$

Thus, all transition in $A_i$ is replaced in $A'_i$ by $m+1$ transitions in the manner stated above. Also, since this construction preserves the synchronization between components and the new rules introduced does not violate the restriction on transitions of deterministic Watson-Crick automata, the system $A'$ recognizes the same language as $A$.

**Theorem 2:** Any language recognized by a deterministic parallel communicating Watson-Crick automata system of degree 2, with injective complementarity relation, can be also recognized by a 4-head deterministic automaton.

Proof. This proof is also similar to the proof in [6] for parallel communicating Watson-Crick automaton. Here we prove for deterministic parallel communicating Watson-Crick automaton. Let $A = (V, \rho, A_1, A_2, K)$ be a deterministic parallel communicating Watson-Crick automata system of degree 2, accepting the language $L \subseteq V^*$, where $A_1 = (V, \rho, Q_1, q_1, F_1, \delta_1)$, $A_2 = (V, \rho, Q_2, q_2, F_2, \delta_2)$, and $K = \{K_1, K_2\}$. Since the relation $\rho$ is injective, we take $\rho$ as the identity relation; thus all components have on both tapes the same word $w \in V^*$. Also, we suppose that in every component only rules of the form $q_i\binom{w_1}{w_2} \to q_j$ with $|w_1 w_2| \leq 1$ are present. The 4-head deterministic finite automaton which accepts the same language as A is constructed in the following manner.

Let us construct now a 4-head automaton $M = (Q, V, \delta, q_0, F)$ where $Q = Q_1 \times Q_2$, $q_0 = (q_1, q_2)$, $F = F_1 \times F_2$, and the transition function $\delta$ is as follows:
$\delta((p, q), w_1, w_2, w_3, w_4) = (p_1, q_1)$ whenever $p, q \notin K$, $p\binom{w_1}{w_2} \to p_1$ is in $\delta_1$ and $q\binom{w_1}{w_2} \to q_1$ is in $\delta_2$.
As both $A_1$ and $A_2$ are deterministic therefore they are is only one rule in $A_1$ which in state p reads $w_1$ in the upper strand and $w_2$ in the lower strand, similar situation holds for $A_2$ also, therefore the finite automaton on reading $w_1$, $w_2$, $w_3$ and $w_4$ in state (p,q) goes only to state $(p_1, q_1)$ and no other transition is present in M which reads $w_1$, $w_2$, $w_3$ and $w_4$ in state (p,q) and goes to some other state.

$\delta((K_2, q), \lambda, \lambda, \lambda, \lambda) = (q, q)$;
$\delta((p, K_2), \lambda, \lambda, \lambda, \lambda) = (p, p)$.

At any step the automaton M simulates the moves of the two components of A. If the components are not in a query state and they read $w_1$, $w_2$, $w_3$ and $w_4$ from the input tape then in M each head reads $w_1$, $w_2$, $w_3$ and $w_4$, respectively, and it enters into the corresponding state which belongs to $Q_1 \times Q_2$ and thus holds the current state of both the components. Otherwise, if M is in state $(K_2, q)$ or $(q, K_1)$, it just simulate the query by entering state (q, q) and leaving the input unchanged. Since a word is accepted by M only if it is in a final state and all the reading heads have finished parsing the input, then $w \in L(A)$ implies $w \in L(M)$ and hence $L(A) \subseteq L(M)$.

Let now w be a word accepted by M. From the construction of the transition function $\delta$, each step in M can be translated into a step in A when the input is of the form $\begin{bmatrix}w\\w\end{bmatrix}$. Moreover as M accepts w all 4 heads of M have completely read the input and the automaton is in a final state. This implies that at the same step both components of system A are in final states and have completely parsed the input. So, we have $w \in L(A)$ and hence $L(M) \subseteq L(A)$.

An inspection of the transition rules introduced in M shows that it is deterministic.

**Theorem 3:** Any language recognized by a deterministic parallel communicating Watson-Crick automata system of degree n, with injective complementarity relation, can be also recognized by a 2n-head deterministic automaton.

Proof: The proof is an extension of the proof in Theorem 8. We form the 2n head multihead deterministic finite automaton in a similar manner as in Theorem 8.

**Example 1**

Let $A = (\{a,b,c,\#\}, \rho, A_1, A_2, A_3, K)$ be a parallel communicating Watson-Crick automata system which accepts the language $L = \{a^{n^2+1}, \text{ where n is even and } n > 1\}$ where $\rho = \{(a,b), (a,c), (a,\#)\}$, $K = \{K_1, K_2, K_3\}$ and $Q = \{q_0, q_1, q_2, q_3, q_4, q_5, q_6\}$

The components of A are as follows:

$A_1$ = ( {a,b,c,#}, K∪Q∪{($q_0$,b), ($q_0$,c), ($q_1$,λ), ($q_2$,λ), ($q_3$,c), ($q_3$,b), ($q_3$,#), ($q_4$,b), ($q_4$,c), ($q_5$,λ), $s_2$, $s_3$} , $q_0$, {$q_6$}, $\delta_1$),

$A_2$ = ( {a,b,c,#}, K∪Q∪{($q_0$,b), ($q_0$,c), ($q_1$,λ), ($q_2$,λ), ($q_3$,c), ($q_3$,b), ($q_3$,#), ($q_4$,b), ($q_4$,c), ($q_5$,λ), ($q_0$,b,λ), ($q_0$,c,λ), ($q_1$,λ,b), ($q_2$,λ,λ), ($q_3$,c,b), ($q_3$,b,c), ($q_3$,#,c), ($q_4$,b,c), ($q_4$,c,λ), ($q_5$,λ,c), ($q_5$,λ,#), $s_3$} , $q_0$, {$q_6$}, $\delta_2$),

and

$A_3$ = ( {a,b,c,#}, K∪Q∪{($q_0$,b,λ), ($q_0$,c,λ), ($q_1$,λ,b), ($q_2$,λ,λ), ($q_3$,c,b), ($q_3$,b,c), ($q_3$,#,c), ($q_4$,b,c), ($q_4$,c,λ), ($q_5$,λ,c), ($q_5$,λ,#), ($q_0$,b,λ,λ), ($q_0$,c,λ,x) x∈{b,c}, ($q_1$,λ,b,x) x∈{b,c}, ($q_2$,λ,λ,x) x∈{b,c}, ($q_3$,c,b,x) x∈{b,c}, ($q_3$,b,c,x) x∈{b,c}, ($q_3$,#,c,#), ($q_4$,b,c,x) x∈{b,c}, ($q_4$,c,λ,x) x∈{b,c}, ($q_5$,λ,c,λ), ($q_5$,λ,#,λ), $p_1$} , $q_0$, {$q_6$}, $\delta_3$).

The transition functions of the three components of A are defined in Table 1.

Table 1: Transition function of components of A

| Component $A_1$ | Component $A_2$ | Component $A_3$ |
|---|---|---|
| $\delta_1(s_2,\binom{\lambda}{\lambda})=s_3$ | $\delta_2(q,\binom{\lambda}{\lambda})=K_1$ for all q∈Q | $\delta_3(q,\binom{\lambda}{\lambda})=p_1$ for all q∈Q |
| $\delta_1(s_3,\binom{\lambda}{\lambda})=K_3$ | $\delta_2(s_3,\binom{\lambda}{\lambda})=K_3$ | $\delta_3(p_1,\binom{\lambda}{\lambda})=K_2$ |
| $\delta_1(q_0,\binom{a}{b})=(q_0,b)$ | $\delta_2((q_0,b),\binom{\lambda}{\lambda})=(q_0,b,\lambda)$ | $\delta_3((q_0,b,\lambda),\binom{\lambda}{\lambda})=(q_0,b,\lambda,\lambda)$ |
| $\delta_1((q_0,b),\binom{\lambda}{\lambda})=s_2$ | $\delta_2((q_0,b,\lambda),\binom{\lambda}{\lambda})=s_3$ | $\delta_3((q_0,b,\lambda,\lambda),\binom{\lambda}{\lambda})=q_0$ |
| $\delta_1(q_0,\binom{a}{c})=(q_0,c)$ | $\delta_2((q_0,c),\binom{\lambda}{\lambda})=(q_0,c,\lambda)$ | $\delta_3((q_0,c,\lambda),\binom{a}{x})=(q_0,c,\lambda,x)$ x∈{b,c} |
| $\delta_1((q_0,c),\binom{\lambda}{\lambda})=s_2$ | $\delta_2((q_0,c,\lambda),\binom{\lambda}{\lambda})=s_3$ | $\delta_3((q_0,c,\lambda,x),\binom{\lambda}{\lambda})=q_1$ x∈{b,c} |
| $\delta_1(q_1,\binom{\lambda}{\lambda})=(q_1,\lambda)$ | $\delta_2((q_1,\lambda),\binom{a}{b})=(q_1,\lambda,b)$ | $\delta_3((q_1,\lambda,b),\binom{a}{x})=(q_1,\lambda,b,x)$ x∈{b,c} |
| $\delta_1((q_1,\lambda),\binom{\lambda}{\lambda})=s_2$ | $\delta_2((q_1,\lambda,b),\binom{\lambda}{\lambda})=s_3$ | $\delta_3((q_1,\lambda,b,x),\binom{\lambda}{\lambda})=q_2$ x∈{b,c} |
| $\delta_1(q_2,\binom{\lambda}{\lambda})=(q_2,\lambda)$ | $\delta_2((q_2,\lambda),\binom{\lambda}{\lambda})=(q_2,\lambda,\lambda)$ | $\delta_3((q_2,\lambda,\lambda),\binom{a}{x})=(q_2,\lambda,\lambda,x)$ x∈{b,c} |
| $\delta_1((q_2,\lambda),\binom{\lambda}{\lambda})=s_2$ | $\delta_2((q_2,\lambda,\lambda),\binom{\lambda}{\lambda})=s_3$ | $\delta_3((q_2,\lambda,\lambda,x),\binom{\lambda}{\lambda})=q_3$ x∈{b,c} |
| $\delta_1(q_3,\binom{a}{c})=(q_3,c)$ | $\delta_2((q_3,c),\binom{a}{b})=(q_3,c,b)$ | $\delta_3((q_3,c,b),\binom{a}{x})=(q_3,c,b,x)$ x∈{b,c} |
| $\delta_1((q_3,c),\binom{\lambda}{\lambda})=s_2$ | $\delta_2((q_3,c,b),\binom{\lambda}{\lambda})=s_3$ | $\delta_3((q_3,c,b,x),\binom{\lambda}{\lambda})=q_3$ x∈{b,c} |
| $\delta_1(q_3,\binom{a}{b})=(q_3,b)$ | $\delta_2((q_3,b),\binom{a}{c})=(q_3,b,c)$ | $\delta_3((q_3,b,c),\binom{a}{x})=(q_3,b,c,x)$ x∈{b,c} |
| $\delta_1((q_3,b),\binom{\lambda}{\lambda})=s_2$ | $\delta_2((q_3,b,c),\binom{\lambda}{\lambda})=s_3$ | $\delta_3((q_3,b,c,x),\binom{\lambda}{\lambda})=q_4$ x∈{b,c} |
| $\delta_1(q_3,\binom{a}{\#})=(q_3,\#)$ | $\delta_2((q_3,\#),\binom{a}{c})=(q_3,\#,c)$ | $\delta_3((q_3,\#,c),\binom{a}{\#})=(q_3,\#,c,\#)$ |
| $\delta_1((q_3,\#),\binom{\lambda}{\lambda})=s_2$ | $\delta_2((q_3,\#,c),\binom{\lambda}{\lambda})=s_3$ | $\delta_3((q_3,\#,c,\#),\binom{\lambda}{\lambda})=q_5$ |
| $\delta_1(q_4,\binom{a}{b})=(q_4,b)$ | $\delta_2((q_4,b),\binom{a}{c})=(q_4,b,c)$ | $\delta_3((q_4,b,c),\binom{a}{x})=(q_4,b,c,x)$ x∈{b,c} |
| $\delta_1((q_4,b),\binom{\lambda}{\lambda})=s_2$ | $\delta_2((q_4,b,c),\binom{\lambda}{\lambda})=s_3$ | $\delta_3((q_4,b,c,x),\binom{\lambda}{\lambda})=q_4$ |
| $\delta_1(q_4,\binom{a}{c})=(q_4,c)$ | $\delta_2((q_4,c),\binom{\lambda}{\lambda})=(q_4,c,\lambda)$ | $\delta_3((q_4,c,\lambda),\binom{a}{x})=(q_4,c,\lambda,x)$ x∈{b,c} |
| $\delta_1((q_4,c),\binom{\lambda}{\lambda})=s_2$ | $\delta_2((q_4,c,\lambda),\binom{\lambda}{\lambda})=s_3$ | $\delta_3((q_4,c,\lambda,x),\binom{\lambda}{\lambda})=q_1$ x∈{b,c} |
| $\delta_1(q_5,\binom{\lambda}{\lambda})=(q_5,\lambda)$ | $\delta_2((q_5,\lambda),\binom{a}{c})=(q_5,\lambda,c)$ | $\delta_3((q_5,\lambda,c),\binom{\lambda}{\lambda})=(q_5,\lambda,c,\lambda)$ |
| $\delta_1((q_5,\lambda),\binom{\lambda}{\lambda})=s_2$ | $\delta_2((q_5,\lambda,c),\binom{\lambda}{\lambda})=s_3$ | $\delta_3((q_5,\lambda,c,\lambda),\binom{\lambda}{\lambda})=q_5$ |
| | $\delta_2((q_5,\lambda),\binom{a}{\#})=(q_5,\lambda,\#)$ | $\delta_3((q_5,\lambda,\#),\binom{\lambda}{\lambda})=(q_5,\lambda,\#,\lambda)$ |
| | $\delta_2((q_5,\lambda,\#),\binom{\lambda}{\lambda})=s_3$ | $\delta_3((q_5,\lambda,\#,\lambda),\binom{\lambda}{\lambda})=q_6$ |

From Table 1, we see that that the Watson-Crick automaton in each component is deterministic in nature thus the above mentioned parallel communicating system is a deterministic parallel communicating Watson-Crick automata system with non-injective complementarity relation.

If we consider a uniletter string having the form $a^{n^2+1}$ where n is even and n>1. Then one of the many complementarity strings for such a uniletter string is of the form $b^n c^n b^n c^n$........$b^n c^n$# where number of times $b^n c^n$ pair is repeated is n/2 and where complementarity relation ρ = {( $a$,b), ($a$,c), ($a$,#)}. But if the uniletter string is not of the form $a^{n^2+1}$ then we will never get a complementarity string of the form $b^n c^n b^n c^n$........$b^n c^n$# where number of times $b^n c^n$ pair is repeated is n/2, thus if we check for such a complementarity string and we get such a complementarity string we know that the upper strand has a uniletter string of the form $a^{n^2+1}$ .

The above system does the checking for such a complementarity string in the following manner the first and second component checks whether the complementarity string in the lower strand is of the form $b^n c^n b^n c^n$........$b^n c^n$# and the first and third component checks whether the number of such $b^n c^n$ pair is n/2. If the complementarity string is not of the form $b^n c^n b^n c^n$........$b^n c^n$# then the first and second component will not reach its final state and if the number of $b^n c^n$ pair is not n/2, then the first and third component will not reach its final state. All the components reach their respective final states and at the same time only when the complementarity string is of the form $b^n c^n b^n c^n$........$b^n c^n$# where number of times $b^n c^n$ pair is repeated is n/2 and we can get such a complementarity string only if the upper strand is of the form $a^{n^2+1}$ where n is even and n>1 . Thus the above mentioned systems accepts the non-regular uniletter language $a^{n^2+1}$ where n is even and n>1. A detailed explanation on how we obtained the above mentioned system is in Appendix 1.

**Example 2:** L = {#$w_1$*$x_1$........#$w_n$*$x_n$\$|n≥0, $w_i$ ∈{a,b}$^*$, $x_i$∈ {$a,b$}$^*$, ∃i∃j :$w_i$=$w_j$, $x_i$≠$x_j$} is accepted by a deterministic Watson-Crick automaton with non-injective complementarity relation.

Let, M=(V,ρ,Q,$q_0$,F,δ) be a Watson-Crick automaton,

where V={a,b,$v_{m1}$,$v_{m2}$,#,*},ρ={(a,a),(#,#),(#,$v_{m1}$),(#,$v_{m2}$),(b,b),(*,*),(\$,\$)}, Q ={$q_0$, $q_l$, $q_w$, $q_x$, $q_{lf}$, $q_{uf}$, $q_f$},F={ $q_f$},and we have the following transitions:

$q_0\binom{\#}{\#}$→$q_0$, $q_0\binom{a}{a}$→$q_0$, $q_0\binom{b}{b}$→$q_0$, $q_0\binom{*}{*}$→$q_0$, $q_0\binom{\#}{v_{m1}}$→$q_l$, $q_l\binom{\lambda}{a}$→$q_l$, $q_l\binom{\lambda}{b}$→$q_l$, $q_l\binom{\lambda}{*}$→$q_l$, $q_l\binom{\lambda}{\#}$→$q_l$, $q_l\binom{\lambda}{v_{m2}}$→$q_w$, $q_w\binom{*}{*}$→$q_x$, $q_w\binom{a}{a}$→$q_w$, $q_w\binom{b}{b}$→$q_w$, $q_x\binom{a}{a}$→$q_x$, $q_x\binom{b}{b}$→$q_x$, $q_x\binom{a}{b}$→$q_{lf}$, $q_x\binom{b}{a}$→$q_{lf}$, $q_{lf}\binom{\lambda}{x}$→$q_{lf}$ x∈{a,b,*,#}, $q_{lf}\binom{\lambda}{\$}$→$q_{uf}$, $q_{uf}\binom{x}{\lambda}$→$q_{uf}$, $q_{uf}\binom{\$}{\lambda}$→$q_f$.

L is not accepted by any k-head deterministic finite automaton [8]. In Example 5, L is accepted by deterministic Watson-Crick automaton by using its non-injective complementarity relation property. '$v_{m1}$' and '$v_{m2}$' are used as complements of '#' to guess the two words in the input string which have their w parts equal but x parts not equal.

Then the two guessed words are compared and if they don't match at any position in their "x" parts but match in their "w" parts then the input string is accepted. If there is no two words in the input string such that there "w" parts are equal and "x" parts are not then no matter where '$v_{m1}$' and '$v_{m2}$' are placed in place of '#' it will never be accepted.

**Theorem 4:** A deterministic parallel communicating Watson-Crick automata system can accept some non-regular uniletter language.

Proof: The proof follows directly from Example 1.

**Theorem 5:** $L_{SDPCWKS}$-$L_{DWK}$≠ ∅ where $L_{DPCWKS}$ is the set of languages accepted by deterministic parallel communicating Watson-Crick automata systems and $L_{DWK}$ is the set of languages accepted by non-deterministic Watson-Crick automata.

Proof: The computational power of strongly deterministic parallel communicating Watson-Crick automata systems is same as that of a deterministic multihead finite automata so, strongly deterministic parallel communicating Watson-Crick automata system can accept the language L={$w_1$#$w_2$#$w_3$#$w_4$#$w_5$#$w_6$, where $w_1$=$w_6$, $w_2$=$w_5$, $w_3$=$w_4$ and $w_1,w_2,w_3,w_4,w_5,w_6$∈{a,b}$^*$} which cannot be accepted by any multihead finite automata with two heads. Thus it cannot be accepted by any non-deterministic Watson-Crick automaton and hence cannot be accepted by any deterministic Watson-Crick automata.

**Theorem 6:** $L_{DWK}$-$L_{SDPCWKS}$≠ ∅ where $L_{DPCWKS}$ is the set of languages accepted by deterministic parallel communicating Watson-Crick automata systems and $L_{DWK}$ is the set of languages accepted by non-deterministic Watson-Crick automata.

Proof: Deterministic multihead finite automata cannot accept the language L = {#$w_1$*$x_1$........#$w_n$*$x_n$\$|n≥0, $w_i$ ∈{a,b}$^*$, $x_i$∈ {$a,b$}$^*$, ∃i∃j :$w_i$=$w_j$, $x_i$≠$x_j$}, hence there is no strongly deterministic parallel communicating Watson-Crick automata systems that can accept L. In Example 2, we show that a deterministic Watson-Crick automaton with non-injective complementarity relation can accept L, which completes the proof.

**Theorem 7:** Strongly deterministic parallel communicating Watson-Crick automata systems and deterministic Watson-Crick automata are incomparable.

Proof: Proof follows directly from the Theorem 5 and 6.

**Theorem 8:** Strongly deterministic parallel communicating Watson-Crick automata systems and non-deterministic Watson-Crick automata are incomparable.

Proof: Proof follows directly from the proofs of Theorem 5 and 6.

**Theorem 9:** Parallel communicating Watson-Crick automata systems with injective complementarity relation cannot accept uniletter non-regular languages.

The proof of this Theorem is in [7].

**Theorem 10:** Non-deterministic multihead finite automata cannot accept uniletter non-regular languages.

The proof of this Theorem is in [7].

**Theorem 11:** $L_{DPCWKS}$-$L_{PCWKS \text{ with injective complementarity relation}} \neq \emptyset$ where $L_{DPCWKS}$ is the set of languages accepted by deterministic parallel communicating Watson-Crick automata systems and $L_{PCWKS \text{ with injective complementarity relation}}$ is the set of languages accepted by non-deterministic parallel communicating Watson-Crick automata systems with injective complementarity relation.

Proof: From Theorem 9, we know that no non-regular uniletter language is accepted by any non-deterministic parallel communicating Watson-Crick automata system with injective complementarity relation. In Example 1, we see that deterministic parallel communicating Watson-Crick automata with non-injective complementarity relation can accept a non-regular uniletter language, which proves the theorem.

**Theorem 12:** $L_{DPCWKS}$-$L_{NFA(K)} \neq \emptyset$ where $L_{DPCWKS}$ is the set of languages accepted by deterministic parallel communicating Watson-Crick automata systems and $L_{NFA(K)}$ is the set of languages accepted by non-deterministic multihead finite automata.

Proof: From Theorem 10, we know that no non-regular uniletter language is accepted by any non-deterministic multihead finite automata. In Example 1, we see that deterministic parallel communicating Watson-Crick automata with non-injective complementarity relation can accept a non-regular uniletter language, which proves the theorem.

**Theorem 13:** Any language recognized by a deterministic parallel communicating finite automata system can be also recognized by a multihead deterministic finite automaton.

Proof of the above stated Theorem is in [1].

**Theorem 14:** Any language recognized by a deterministic parallel communicating finite automata system of degree n can be also recognized by strongly deterministic parallel communicating Watson-Crick automata system of degree n.

Proof: The proof follows from the definitions of deterministic parallel communicating finite automata system and strongly deterministic parallel communicating Watson-Crick automata system.

**Theorem 15:** Any language recognized by deterministic multihead finite automata can be also recognized by strongly deterministic parallel communicating Watson-Crick automata system.

Proof: Proof follows from Theorem 14 and Theorem 13.

**Theorem 16:** Strongly deterministic parallel communicating Watson-Crick automata systems and deterministic multihead finite automata have the same computational powers.

Proof: Proof follows from Theorem 1 and Theorem 16.

**Theorem 17:** Strongly deterministic parallel communicating Watson-Crick automata systems are proper subset of deterministic parallel communicating Watson-Crick automata systems.

Proof: The definitions of strongly deterministic parallel communicating Watson-Crick automata systems and deterministic parallel communicating Watson-Crick automata systems shows that for every strongly deterministic parallel communicating Watson-Crick automata system there exists a deterministic parallel communicating Watson-Crick automata system which accepts the same language. Now the computational power of strongly deterministic parallel communicating Watson-Crick automata systems is same as that of deterministic multihead finite automata and thus, strongly deterministic parallel communicating Watson-Crick automata system cannot accept L={#$w_1$*$x_1$.........#$w_n$*$x_n$\$|n≥0, $w_i \in \{a,b\}^*$, $x_i \in \{a,b\}^*$, $\exists i \exists j : w_i = w_j$, $x_i \neq x_j$} which is accepted by a deterministic Watson-Crick automaton as shown in Example 2. As a deterministic Watson-Crick automaton can be considered as a deterministic parallel communicating Watson-Crick automata system of degree 1, so deterministic parallel communicating Watson-Crick automata system accept a language not accepted by any strongly deterministic parallel communicating Watson-Crick automata system. Hence strongly deterministic parallel communicating Watson-Crick automata systems are a proper subset of deterministic parallel communicating Watson-Crick automata systems.

VIII. CONCLUSION

In this paper, we have introduced deterministic variant of parallel communicating Watson-Crick automata systems. We show that the deterministic variant similar to the non-deterministic variant can accept non-regular uniletter language using non-injective complementarity relation. We compare the computational complexity of such a model with multihead finite automata and parallel communicating finite automata. We further establish that strongly deterministic parallel communicating Watson-Crick automata systems are a proper subset of deterministic parallel communicating Watson-Crick automata systems. Moreover, we also show that strongly deterministic parallel communicating Watson-Crick automata systems are incomparable to both non-deterministic Watson-Crick automata and deterministic Watson-Crick automata in terms of computational power.

# Appendix 1

Consider a multihead deterministic finite automaton with 3 heads $M=(3,V,Q,q_0,F,\delta)$ where $V=\{b,c,\#\}$, $Q=\{q_0, q_1, q_2, q_3, q_4, q_5, q_6\}$, $F=\{q_6\}$ which accept the language $L=\{b^n c^n b^n c^n ..... b^n c^n \# |$ where number of such $b^n c^n$ pairs is $n/2$, n is even and $n>1\}$

The transitions of M are as follows:

$\delta(q_0,b,\lambda,\lambda)=q_0$, $\delta(q_0,c,\lambda,x)=q_1$ $x\in\{b,c\}$, $\delta(q_1,\lambda,b,x)=q_2$ $x\in\{b,c\}$, $\delta(q_2,\lambda,\lambda,x)=q_3$ $x\in\{b,c\}$, $\delta(q_3,c,b,x)=q_3$ $x\in\{b,c\}$, $\delta(q_3,b,c,x)=q_4$ $x\in\{b,c\}$, $\delta(q_3,\#,c,\#)=q_5$, $\delta(q_4,b,c,x)=q_4$ $x\in\{b,c\}$, $\delta(q_4,c,\lambda,x)=q_1$ $x\in\{b,c\}$, $\delta(q_5,\lambda,c,\lambda)=q_5$, $\delta(q_5,\lambda,\#,\lambda)=q_6$.

The above automaton M works as follows the first two heads check whether the input is of the form $b^n c^n b^n c^n ..... b^n c^n \#$ and the first and the third head checks whether the number of such $b^n c^n$ pairs is $n/2$ where n is even and $n>1$.

Now $A'=(V,A'_1,A'_2,A'_3,K)$ a parallel communicating deterministic finite automata system is derived from M using the transformation rules stated in [1] to obtain a parallel communicating deterministic finite automata system which accepts the same language as a deterministic multihead finite automaton. Thus A' accepts the same language as M. $V=\{b,c,\#\}$, $K=\{K_1,K_2,K_3\}$

The components of A' are as follows:

$A'_1 = (\{b,c,\#\}, K\cup Q\cup\{(q_0,b), (q_0,c), (q_1,\lambda), (q_2,\lambda), (q_3,c), (q_3,b), (q_3,\#), (q_4,b), (q_4,c), (q_5,\lambda), s_2, s_3\}, q_0, \{q_6\}, \delta'_1)$,

$A'_2 = (\{b,c,\#\}, K\cup Q\cup\{(q_0,b), (q_0,c), (q_1,\lambda), (q_2,\lambda), (q_3,c), (q_3,b), (q_3,\#), (q_4,b), (q_4,c), (q_5,\lambda), (q_0,b,\lambda), (q_0,c,\lambda), (q_1,\lambda,b), (q_2,\lambda,\lambda), (q_3,c,b), (q_3,b,c), (q_3,\#,c), (q_4,b,c), (q_4,c,\lambda), (q_5,\lambda,c), (q_5,\lambda,\#), s_3\}, q_0, \{q_6\}, \delta'_2)$,

and

A'$_3$ = ( {b,c,#}, K∪Q∪{(q$_0$,b,λ), (q$_0$,c,λ), (q$_1$,λ,b), (q$_2$,λ,λ), (q$_3$,c,b), (q$_3$,b,c), (q$_3$,#,c), (q$_4$,b,c), (q$_4$,c,λ), (q$_5$,λ,c), (q$_5$,λ,#), (q$_0$,b,λ,λ), (q$_0$,c,λ,x) x∈{b,c}, (q$_1$,λ,b,x) x∈{b,c}, (q$_2$,λ,λ,x) x∈{b,c}, (q$_3$,c,b,x) x∈{b,c}, (q$_3$,b,c,x) x∈{b,c}, (q$_3$,#,c,#), (q$_4$,b,c,x) x∈{b,c}, (q$_4$,c,λ,x) x∈{b,c}, (q$_5$,λ,c,λ), (q$_5$,λ,#,λ),  p$_1$ } , q$_0$, {q$_6$}, δ'$_3$).

The transition functions of the three components of A' are defined in Table 2.

Table 2: Transition function of components of A'

| Component A$_1$ | Component A$_2$ | Component A$_3$ |
|---|---|---|
| δ'$_1$(s$_2$,λ)=s$_3$ | δ'$_2$(q,λ)=K$_1$ for all q∈Q | δ'$_3$(q,λ)=p$_1$ for all q∈Q |
| δ'$_1$(s$_3$,λ)=K$_3$ | δ'$_2$(s$_3$,λ)=K$_3$ | δ'$_3$(p$_1$,λ)=K$_2$ |
| δ'$_1$(q$_0$,b)=(q$_0$,b) | δ'$_2$((q$_0$,b),λ)=(q$_0$,b,λ) | δ'$_3$((q$_0$,b,λ),λ)=(q$_0$,b,λ,λ) |
| δ'$_1$((q$_0$,b),λ)=s$_2$ | δ'$_2$((q$_0$,b,λ),λ)=s$_3$ | δ'$_3$((q$_0$,b,λ,λ),λ)=q$_0$ |
| δ'$_1$(q$_0$,c)=(q$_0$,c) | δ'$_2$((q$_0$,c),λ)=(q$_0$,c,λ) | δ'$_3$((q$_0$,c,λ),x)=(q$_0$,c,λ,x) x∈{b,c} |
| δ'$_1$((q$_0$,c),λ)=s$_2$ | δ'$_2$((q$_0$,c,λ),λ)=s$_3$ | δ'$_3$((q$_0$,c,λ,x),λ)=q$_1$ x∈{b,c} |
| δ'$_1$(q$_1$,λ)=(q$_1$,λ) | δ'$_2$((q$_1$,λ),b)=(q$_1$,λ,b) | δ'$_3$((q$_1$,λ,b),x)=(q$_1$,λ,b,x) x∈{b,c} |
| δ'$_1$((q$_1$,λ),λ)=s$_2$ | δ'$_2$((q$_1$,λ,b),λ)=s$_3$ | δ'$_3$((q$_1$,λ,b,x),λ)=q$_2$ x∈{b,c} |
| δ'$_1$(q$_2$,λ)=(q$_2$,λ) | δ'$_2$((q$_2$,λ),λ)=(q$_2$,λ,λ) | δ'$_3$((q$_2$,λ,λ),x)=(q$_2$,λ,λ,x) x∈{b,c} |
| δ'$_1$((q$_2$,λ),λ)=s$_2$ | δ'$_2$((q$_2$,λ,λ),λ)=s$_3$ | δ'$_3$((q$_2$,λ,λ,x),λ)=q$_3$ x∈{b,c} |
| δ'$_1$(q$_3$,c)=(q$_3$,c) | δ'$_2$((q$_3$,c),b)=(q$_3$,c,b) | δ'$_3$((q$_3$,c,b),x)=(q$_3$,c,b,x) x∈{b,c} |
| δ'$_1$((q$_3$,c),λ)=s$_2$ | δ'$_2$((q$_3$,c,b),λ)=s$_3$ | δ'$_3$((q$_3$,c,b,x),λ)=q$_3$ x∈{b,c} |
| δ'$_1$(q$_3$,b)=(q$_3$,b) | δ'$_2$((q$_3$,b),c)=(q$_3$,b,c) | δ'$_3$((q$_3$,b,c),x)=(q$_3$,b,c,x) x∈{b,c} |
| δ'$_1$((q$_3$,b),λ)=s$_2$ | δ'$_2$((q$_3$,b,c),λ)=s$_3$ | δ'$_3$((q$_3$,b,c,x),λ)=q$_4$ x∈{b,c} |
| δ'$_1$(q$_3$,#)=(q$_3$,#) | δ'$_2$((q$_3$,#),c)=(q$_3$,#,c) | δ'$_3$((q$_3$,#,c),#)=(q$_3$,#,c,#) |
| δ'$_1$((q$_3$,#),λ)=s$_2$ | δ'$_2$((q$_3$,#,c),λ)=s$_3$ | δ'$_3$((q$_3$,#,c,#),λ)=q$_5$ |
| δ'$_1$(q$_4$,b)=(q$_4$,b) | δ'$_2$((q$_4$,b),c)=(q$_4$,b,c) | δ'$_3$((q$_4$,b,c),x)=(q$_4$,b,c,x) x∈{b,c} |
| δ'$_1$((q$_4$,b),λ)=s$_2$ | δ'$_2$((q$_4$,b,c),λ)=s$_3$ | δ'$_3$((q$_4$,b,c,x),λ)=q$_4$ |
| δ'$_1$(q$_4$,c)=(q$_4$,c) | δ'$_2$((q$_4$,c),λ)=(q$_4$,c,λ) | δ'$_3$((q$_4$,c,λ),x)=(q$_4$,c,λ,x) x∈{b,c} |
| δ'$_1$((q$_4$,c),λ)=s$_2$ | δ'$_2$((q$_4$,c,λ),λ)=s$_3$ | δ'$_3$((q$_4$,c,λ,x),λ)=q$_1$ x∈{b,c} |
| δ'$_1$(q$_5$,λ)=(q$_5$,λ) | δ'$_2$((q$_5$,λ),c)=(q$_5$,λ,c) | δ'$_3$((q$_5$,λ,c),λ)=(q$_5$,λ,c,λ) |
| δ'$_1$((q$_5$,λ),λ)=s$_2$ | δ'$_2$((q$_5$,λ,c),λ)=s$_3$ | δ'$_3$((q$_5$,λ,c,λ),λ)=q$_5$ |
|  | δ'$_2$((q$_5$,λ),#)=(q$_5$,λ,#) | δ'$_3$((q$_5$,λ,#),λ)=(q$_5$,λ,#,λ) |
|  | δ'$_2$((q$_5$,λ,#),λ)=s$_3$ | δ'$_3$((q$_5$,λ,#,λ),λ)=q$_6$ |

Now A=({a,b,c,#},ρ,A$_1$,A$_2$,A$_3$,K) a deterministic parallel communicating Watson-Crick finite automata system is derived from A' in the following manner:

$\rho=\{(a,b), (a,c), (a,\#)\}$ is the complementarity relation of A.

For every component $A'_i$ in A' where $A'_i=(\{b,c,\#\},Q'_i,q_0,F'_i,\delta'_i)$ is a deterministic finite automaton there is a deterministic Watson-Crick automaton $A_i=(\{a,b,c,\#\},\rho,Q'_i,q_0,F'_i,\delta_i)$ in A where $\delta_i$ is obtained from $\delta'_i$ as follows:

For every transition of the form $\delta'_i(q_i,\lambda)=q_j$ in $A'_i$ where $q_i,q_j \in Q'_i$ introduce $\delta_i\left(q_i,\binom{\lambda}{\lambda}\right) = q_j$ in $A_i$.

For every transition of the form $\delta'_i(q_i,y)=q_j$ in $A'_i$ where $q_i,q_j \in Q'_i$ and $y\in\{b,c,\#\}$ introduce $\delta_i\left(q_i,\binom{a}{x}\right) = q_j$ in $A_i$.

The deterministic parallel communicating Watson-Crick finite automata system A is derived from A' described below.

$A=(\{a,b,c,\#\},\rho,A_1,A_2,A_3,\{K_1,K_2,K_3\})$ where $\rho=\{(a,b),(a,c),(a,\#)\}$.

The components of A are as follows:

$A_1 = (\{a,b,c,\#\}, K\cup Q\cup\{(q_0,b), (q_0,c), (q_1,\lambda), (q_2,\lambda), (q_3,c), (q_3,b), (q_3,\#), (q_4,b), (q_4,c), (q_5,\lambda), s_2, s_3\}, q_0, \{q_6\}, \delta_1)$,

$A_2 = (\{a,b,c,\#\}, K\cup Q\cup\{(q_0,b), (q_0,c), (q_1,\lambda), (q_2,\lambda), (q_3,c), (q_3,b), (q_3,\#), (q_4,b), (q_4,c), (q_5,\lambda), (q_0,b,\lambda), (q_0,c,\lambda), (q_1,\lambda,b), (q_2,\lambda,\lambda),$
$(q_3,c,b), (q_3,b,c), (q_3,\#,c), (q_4,b,c), (q_4,c,\lambda), (q_5,\lambda,c), (q_5,\lambda,\#), s_3\}, q_0, \{q_6\}, \delta_2)$,

and

$A_3 = (\{a,b,c,\#\}, K\cup Q\cup\{(q_0,b,\lambda), (q_0,c,\lambda), (q_1,\lambda,b), (q_2,\lambda,\lambda), (q_3,c,b), (q_3,b,c), (q_3,\#,c), (q_4,b,c), (q_4,c,\lambda), (q_5,\lambda,c), (q_5,\lambda,\#),$
$(q_0,b,\lambda,\lambda), (q_0,c,\lambda,x)\ x\in\{b,c\}, (q_1,\lambda,b,x)\ x\in\{b,c\}, (q_2,\lambda,\lambda,x)\ x\in\{b,c\}, (q_3,c,b,x)\ x\in\{b,c\}, (q_3,b,c,x)\ x\in\{b,c\}, (q_3,\#,c,\#), (q_4,b,c,x)$
$x\in\{b,c\}, (q_4,c,\lambda,x)\ x\in\{b,c\}, (q_5,\lambda,c,\lambda), (q_5,\lambda,\#,\lambda),\ p_1\}, q_0, \{q_6\}, \delta_3)$.

The transition functions of the three components of A are defined in Table 3.

Table 3: Transition function of components of A

| *Component $A_1$* | *Component $A_2$* | *Component $A_3$* |
|---|---|---|
| $\delta_1(s_2,\binom{\lambda}{\lambda})=s_3$ | $\delta_2(q,\binom{\lambda}{\lambda})=K_1$ for all $q\in Q$ | $\delta_3(q,\binom{\lambda}{\lambda})=p_1$ for all $q\in Q$ |
| $\delta_1(s_3,\binom{\lambda}{\lambda})=K_3$ | $\delta_2(s_3,\binom{\lambda}{\lambda})=K_3$ | $\delta_3(p_1,\binom{\lambda}{\lambda})=K_2$ |
| $\delta_1(q_0,\binom{a}{b})=(q_0,b)$ | $\delta_2((q_0,b),\binom{\lambda}{\lambda})=(q_0,b,\lambda)$ | $\delta_3((q_0,b,\lambda),\binom{\lambda}{\lambda})=(q_0,b,\lambda,\lambda)$ |
| $\delta_1((q_0,b),\binom{\lambda}{\lambda})=s_2$ | $\delta_2((q_0,b,\lambda),\binom{\lambda}{\lambda})=s_3$ | $\delta_3((q_0,b,\lambda,\lambda),\binom{\lambda}{\lambda})=q_0$ |
| $\delta_1(q_0,\binom{a}{c})=(q_0,c)$ | $\delta_2((q_0,c),\binom{\lambda}{\lambda})=(q_0,c,\lambda)$ | $\delta_3((q_0,c,\lambda),\binom{a}{x})=(q_0,c,\lambda,x)\ x\in\{b,c\}$ |
| $\delta_1((q_0,c),\binom{\lambda}{\lambda})=s_2$ | $\delta_2((q_0,c,\lambda),\binom{\lambda}{\lambda})=s_3$ | $\delta_3((q_0,c,\lambda,x),\binom{\lambda}{\lambda})=q_1\ x\in\{b,c\}$ |
| $\delta_1(q_1,\binom{\lambda}{\lambda})=(q_1,\lambda)$ | $\delta_2((q_1,\lambda),\binom{a}{b})=(q_1,\lambda,b)$ | $\delta_3((q_1,\lambda,b),\binom{a}{x})=(q_1,\lambda,b,x)\ x\in\{b,c\}$ |
| $\delta_1((q_1,\lambda),\binom{\lambda}{\lambda})=s_2$ | $\delta_2((q_1,\lambda,b),\binom{\lambda}{\lambda})=s_3$ | $\delta_3((q_1,\lambda,b,x),\binom{\lambda}{\lambda})=q_2\ x\in\{b,c\}$ |
| $\delta_1(q_2,\binom{\lambda}{\lambda})=(q_2,\lambda)$ | $\delta_2((q_2,\lambda),\binom{\lambda}{\lambda})=(q_2,\lambda,\lambda)$ | $\delta_3((q_2,\lambda,\lambda),\binom{a}{x})=(q_2,\lambda,\lambda,x)\ x\in\{b,c\}$ |
| $\delta_1((q_2,\lambda),\binom{\lambda}{\lambda})=s_2$ | $\delta_2((q_2,\lambda,\lambda),\binom{\lambda}{\lambda})=s_3$ | $\delta_3((q_2,\lambda,\lambda,x),\binom{\lambda}{\lambda})=q_3\ x\in\{b,c\}$ |
| $\delta_1(q_3,\binom{a}{c})=(q_3,c)$ | | $\delta_3((q_3,c,b),\binom{a}{x})=(q_3,c,b,x)\ x\in\{b,c\}$ |
| $\delta_1((q_3,c),\binom{\lambda}{\lambda}))=s_2$ | $\delta_2((q_3,c),\binom{a}{b}))=(q_3,c,b)$ | $\delta_3((q_3,c,b,x),\binom{\lambda}{\lambda}))=q_3\ x\in\{b,c\}$ |
| | $\delta_2((q_3,c,b),\binom{\lambda}{\lambda}))=s_3$ | |
| $\delta_1(q_3,\binom{a}{b})=(q_3,b)$ | | $\delta_3((q_3,b,c),\binom{a}{x}))=(q_3,b,c,x)\ x\in\{b,c\}$ |
| $\delta_1((q_3,b),\binom{\lambda}{\lambda}))=s_2$ | $\delta_2((q_3,b),\binom{a}{c}))=(q_3,b,c)$ | $\delta_3((q_3,b,c,x),\binom{\lambda}{\lambda}))=q_4\ x\in\{b,c\}$ |
| | $\delta_2((q_3,b,c),\binom{\lambda}{\lambda}))=s_3$ | |
| $\delta_1(q_3,\binom{a}{\#})=(q_3,\#)$ | | |
| $\delta_1((q_3,\#),\binom{\lambda}{\lambda}))=s_2$ | | $\delta_3((q_3,\#,c),\binom{a}{\#}))=(q_3,\#,c,\#)$ |
| | $\delta_2((q_3,\#),\binom{a}{c}))=(q_3,\#,c)$ | $\delta_3((q_3,\#,c,\#),\binom{\lambda}{\lambda}))=q_5$ |
| | $\delta_2((q_3,\#,c),\binom{\lambda}{\lambda}))=s_3$ | |
| $\delta_1(q_4,\binom{a}{b})=(q_4,b)$ | | |
| $\delta_1((q_4,b),\binom{\lambda}{\lambda}))=s_2$ | | $\delta_3((q_4,b,c),\binom{a}{x}))=(q_4,b,c,x)\ x\in\{b,c\}$ |
| | $\delta_2((q_4,b),\binom{a}{c}))=(q_4,b,c)$ | $\delta_3((q_4,b,c,x),\binom{\lambda}{\lambda}))=q_4$ |
| | $\delta_2((q_4,b,c),\binom{\lambda}{\lambda}))=s_3$ | |
| $\delta_1(q_4,\binom{a}{c})=(q_4,c)$ | | |
| $\delta_1((q_4,c),\binom{\lambda}{\lambda}))=s_2$ | | $\delta_3((q_4,c,\lambda),\binom{a}{x}))=(q_4,c,\lambda,x)\ x\in\{b,c\}$ |
| | $\delta_2((q_4,c),\binom{\lambda}{\lambda}))=(q_4,c,\lambda)$ | $\delta_3((q_4,c,\lambda,x),\binom{\lambda}{\lambda}))=q_1\ x\in\{b,c\}$ |
| $\delta_1(q_5,\binom{\lambda}{\lambda})=(q_5,\lambda)$ | $\delta_2((q_4,c,\lambda),\binom{\lambda}{\lambda}))=s_3$ | |
| $\delta_1((q_5,\lambda),\binom{\lambda}{\lambda}))=s_2$ | | $\delta_3((q_5,\lambda,c),\binom{\lambda}{\lambda}))=(q_5,\lambda,c,\lambda)$ |
| | $\delta_2((q_5,\lambda),\binom{a}{c}))=(q_5,\lambda,c)$ | $\delta_3((q_5,\lambda,c,\lambda),\binom{\lambda}{\lambda}))=q_5$ |
| | $\delta_2((q_5,\lambda,c),\binom{\lambda}{\lambda}))=s_3$ | |
| | | $\delta_3((q_5,\lambda,\#),\binom{\lambda}{\lambda}))=(q_5,\lambda,\#,\lambda)$ |
| | $\delta_2((q_5,\lambda),\binom{a}{\#}))=(q_5,\lambda,\#)$ | $\delta_3((q_5,\lambda,\#,\lambda),\binom{\lambda}{\lambda}))=q_6$ |
| | $\delta_2((q_5,\lambda,\#),\binom{\lambda}{\lambda}))=s_3$ | |

From the construction of A from A' it is evident that A accepts all those double stranded strings which have a's in its upper strand and the lower strand is of the form $b^nc^nb^nc^n$..... $b^nc^n$# where number of such $b^nc^n$ pairs is n/2, n is even and n>1.

Now consider a string 'w' belonging to the language L=$\{a^{n^2+1},$ where n is even and n > 1$\}$ and suppose the complementarity relation $\rho=\{(a,b), (a,c), (a,\#)\}$. Then one of the many complementarity strings possible for such a 'w' must be of the form $b^nc^nb^nc^n$..... $b^nc^n$# where number of such $b^nc^n$ pairs is n/2, n is even and n>1 and hence 'w' is accepted by A. If 'w' does not belong to L then it can never have a complementarity string which is of the form $b^nc^nb^nc^n$.....$b^nc^n$# where number of such $b^nc^n$ pairs is n/2, n is even and n>1 hence 'w' is not accepted by A. Thus A accepts only those strings which are in L. Thus A accepts L.